# Performance and prospects of polyurethane-based scintillators for neutron and gamma radiation detection

Olga Maiatska*[a], Torsten Dünnebacke[a], Roland Hillebrands[a], Achim Kreuels[a], Martin Kreuels[a], Guntram Pausch[a], Falko Scherwinski[a], Andreas Wolf[a], and Jürgen Stein[a]
[a]Rapiscan Systems GmbH, Heinz-Fangman-Str. 4, D-42287 Wuppertal, Germany

## ABSTRACT

Organic scintillators have been used for decades to detect gamma rays or fast neutrons at low cost. Despite extensive efforts it is still difficult to combine stability, ruggedness, high light yield, and decent pulse-shape discrimination (PSD) in a single scintillator material. We took the unusual approach of embedding phosphors and wavelengths shifters in a solid polyurethane matrix, trying to combine the robustness of a common construction material with scintillating properties and PSD performance. The plastic scintillators resulting from years of research ultimately meet these intentions. Moreover, the production process is less demanding and less complex than that of conventional plastic scintillators. It allows to directly integrate light sensors or electronics into the scintillator. This opens the door for the construction of extremely compact and robust plastic detectors for the simultaneous detection of fast neutrons and gamma rays. This paper presents basic recipes and properties of scintillators based on this novel conception and discusses first applications in neutron imaging and personal radiation protection.

## 1. INTRODUCTION

Organic scintillators, discovered almost sixty years ago [1], are widely used in radiation detection. On the one hand, common plastic scintillators – typically consisting of a plastic matrix comprised with a primary fluor (first additive) and a wavelength shifter (second additive) – can be produced in large sizes and are quite affordable [2]. Their unsurpassed sensitivity in terms of counts per invested dollar makes them attractive for radiation monitoring in many fields, including medicine and homeland security. On the other hand, organic crystals such as stilbene or anthracene and liquid organic scintillators have been the workhorses for fast-neutron detection since pulse-shape discrimination (PSD) was introduced to distinguish neutron (recoil proton) from gamma-ray (recoil electron) signals [3].

So far, the development of plastic scintillators has been focused on polystyrene (PS) or polyvinyl toluene (PVT) based matrices. The production of such conventional plastic scintillators relies on free-radical polymerization under carefully controlled conditions, including oxygen-free processing and thorough purification of the monomeric precursors. The resulting materials are known to be susceptible to ageing effects, as to "fogging" observed in PVT-based materials under harsh environmental conditions [4]. Moreover, they cannot provide reasonable PSD capabilities – except for some recent formulations as Eljen's EJ-276 family[1] that is, however, known to lose performance over time [5].

To overcome these limitations of scintillating plastics, an alternative concept has been explored at Rapiscan Systems GmbH (former Target Systemelektronik GmbH & Co. KG) [6]. This paper reviews basic properties of the polyurethane-(PU-)based scintillators developed since 2018 and discusses first practical applications.

*OMaiatska@rapiscan.com; phone +49 202 76930 20; rapiscansystems.com

[1] https://eljentechnology.com/products/plastic-scintillators/ej-276

# 2. POLYURETHANE-BASED PLASTIC SCINTILLATORS

## 2.1 Approach

Instead of developing new scintillator formulations by using common technologies, a novel concept has been explored and established. The basic idea was to design a transparent, two-component polyurethane matrix that allows to combine advantageous properties of modern plastic scintillators – excellent optical transmission, high refractive index, good light yield and PSD capabilities – with a simplified and highly customizable processing, outstanding mechanical robustness, and long-term environmental stability. To achieve these objectives, a primary scintillation fluor and a wavelength shifter are completely dissolved in the liquid precursor mixture before polymerization. This results in their homogeneous incorporation into the forming polymer network. Both the primary fluor and the wavelength shifter can be varied to meet the desired detector characteristics. While the selection of the primary fluor mostly affects the scintillation light yield, the variation of the wavelength shifter enables a controlled adjustment of the emission spectrum, the scintillation decay time, and the PSD performance. The resulting material is an optically transparent plastic scintillator with a homogeneous distribution of the scintillating components, whose properties can be tailored to meet the specific requirements of different radiation detection applications. The cured material can easily be cut, drilled, milled, and polished without compromising structural integrity or optical quality. The low polymerization shrinkage of $< 1\%$ minimizes internal stresses during curing, ensures dimensional stability of the fabricated scintillators, and allows, at the same time, the direct integration of optical sensors and related electronics into the polymer matrix.

## 2.2 Production process

A primary scintillation fluor (first additive) and a wavelength shifter (second additive) are completely dissolved in the liquid two-component mixture before polymerization. The liquid formulation is carefully degassed to eliminate air bubbles and optical scattering centers, cast into a silicone mold, and cured at about 100°C for about 48 hours. In contrast to the manufacturing of PS- or PVT-based plastic scintillators, neither oxygen-free polymerization nor purification of the monomeric precursors is required. This simplifies the production process and improves its reproducibility.

The mold used for casting can be chosen in accordance with the scintillator's envisaged shape. Complex geometries can thus be produced without complex machining processes. The liquid processing also enables the production of large-volume scintillators that are difficult to manufacture using conventional polymerization techniques.

## 2.3 Basic properties and performance

Over the years, various PU-based scintillator materials have been developed and explored at Rapiscan Systems. M600 was the first one whose performance was comparable to that of conventional PSD-capable plastic materials. It was made available to external groups and characterized in the context of research projects [7][8].

The latest material, M700, performs even better. It is distinguished by a relatively high density (1.164 g/cm$^3$) and refractive index (1.62). The Shore D hardness (82) and softening point (75°C) are comparable with that of Eljen EJ-200[2], a common commercial plastic scintillator. The atomic constituents of M700 are hydrogen (46.7 at%), carbon (43.7 at%), nitrogen (3.6 at%), and oxygen (6.1 at%). The light yield of M700 exceeds that of M600 and equals the light yield of EJ-200 and EJ-276D, while its PSD performance, measured with a Figure of Merit (FOM, see Figure 2), is better than that of EJ-276D, the latest of the commercial plastic materials capable of neutron-gamma discrimination [9]. In contrast to other commercial scintillators, the M700 samples showed no fogging, clouding, yellowing, or other optical degradation in environmental tests including the exposure to elevated temperatures (60°C) and humidity (100%) (Figure 1). Likewise, the light yield and PSD performance remained unchanged [9].

## 2.4 Effect on potential applications

PU-based scintillators generally provide a higher density than common plastic scintillators and comprise not only hydrogen and carbon but also nitrogen and oxygen atoms. This results in a more body-equivalent response to ionizing radiation: The integrated sensor signal (equivalent with the total light produced in the scintillators) in reference gamma radiation fields better approaches the energy dependence of the dose absorbed by a human body. This makes PU-based scintillators good candidates for dose monitoring.

[2] https://eljentechnology.com/products/plastic-scintillators/ej-200-ej-204-ej-208-ej-212

So far, no loss of transparency, decline of light yield, fogging, or other kind of ageing has been observed with PU-based scintillators. As intended, the scintillating materials keep the advantageous mechanical properties of polyurethane: robustness, stress resistance, elasticity, scratch resistance, ease of processing, chemical resistance. This recommends PU-based scintillators for applications where ruggedness and persistence are of critical importance.

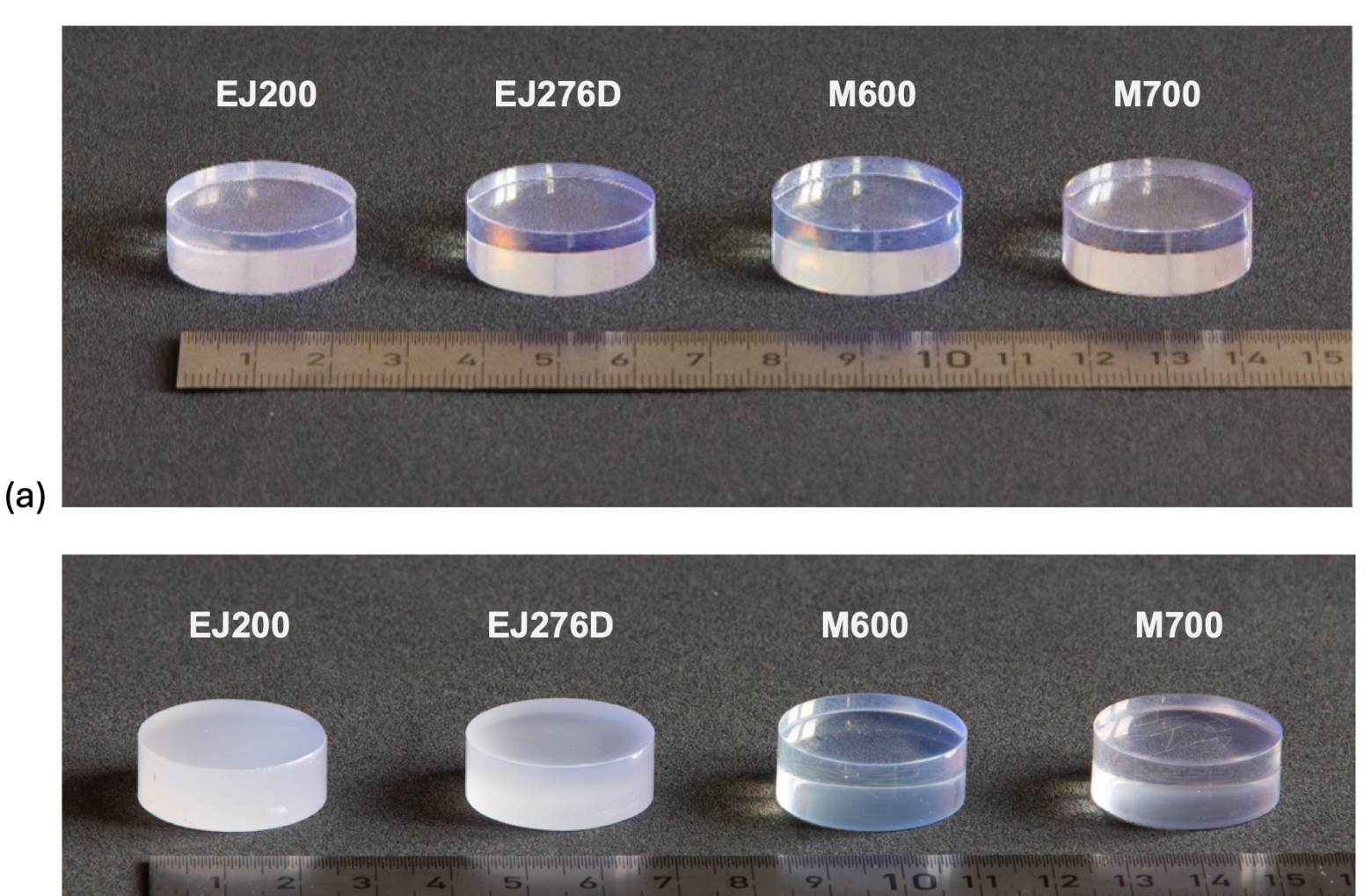

Figure 1. Photographs of various scintillator samples (a) after sample preparation, (b) after 48 hours of “steaming” at 60°C and 48 hours of “freezing” at –18°C. The EJ-200 and EJ-276D samples clearly show degraded optical quality (milky appearance), while the PU-based M600 and M700 samples keep their transparency. The figure was taken from reference [9].

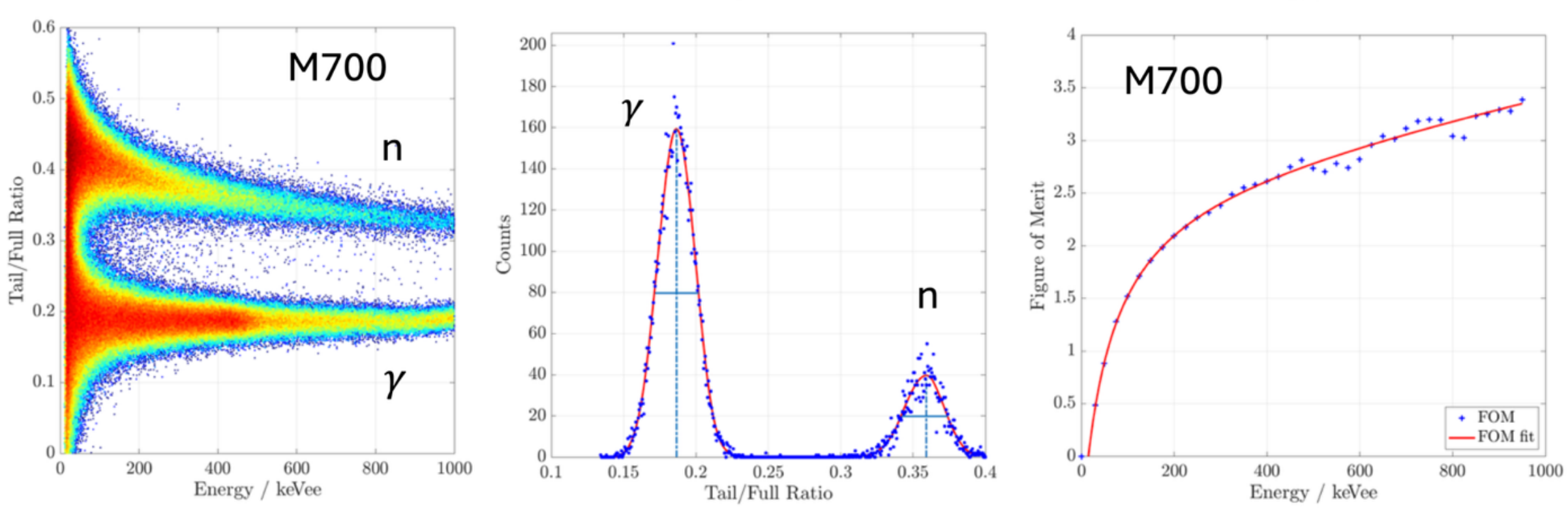

Figure 2. PSD performance of the small, cylindrical (Ø1" × 10 mm) M700 sample shown in Figure 1, read out with a Hamamatsu R10601-100 photomultiplier tube, exposed to neutrons and gamma rays from $^{252}$Cf and $^{137}$Cs sources. The left panels represent the pulse shape parameter $TFR$ (tail-to-full ratio of pulse shapes), plotted versus energy deposition. The $TFR$ spectrum (middle) comprises events of the energy bin 452 to 502 keVee. It visualizes the neutron-gamma separation around the 477 keV Compton edge of $^{137}$Cs, and the fits to the gamma and neutron peaks used to determine a Figure of Merit (FOM). The FOM is defined as the difference of the neutron and gamma peak positions, divided by the sum of both peak widths at half maximum (FWHM) [9]. Corresponding fits performed for multiple energy bins yield the energy dependence of the FOM, shown in the right panel. The figure was taken from reference [9].

Another aspect is the manufacturing process — specifically casting and curing at rather moderate temperatures with negligible polymerization shrinkage — which enables the integration of optical sensors (such as silicon photomultipliers, including their associated carrier boards) into the scintillator material.

# 3. FIRST APPLICATIONS

## 3.1 Dual-particle imaging for proton therapy verification

The properties of PU-based scintillators, especially their competitive light output and PSD performance in combination with the outstanding robustness and longevity, make them a good choice for neutron or dual-particle (neutron and gamma-ray) imaging systems intended for operation outside of laboratory environments. In the NOVO project[3] [10], ruggedness and persistence argue in favor of preferring M700 over the alternative with actually superior performance, namely organic glass scintillators (OGS) [11][12][13]. The project, funded by the European Innovation Council (EIC), aims to develop a robust, modular, clinically deployable detection system named NOVCoDA for simultaneous neutron and gamma-ray imaging in medical cancer treatments with proton beams. Bar-shaped M600 and M700 scintillators fitting with the NOVO design have already been produced [14] (Figure 3). First imaging experiments have been performed with a provisional setup comprising OGS as well as M600 scintillator bars of the same size ($12\times12\times140$ mm$^3$), read out with Hamamatsu S14161-3050HS-04 silicon photomultiplier (SiPM) arrays on both sides [15]. The results confirm the detector concept as well as the usability of M600 for the NOVCoDA array. The M700 bars promising even better results are just being tested.

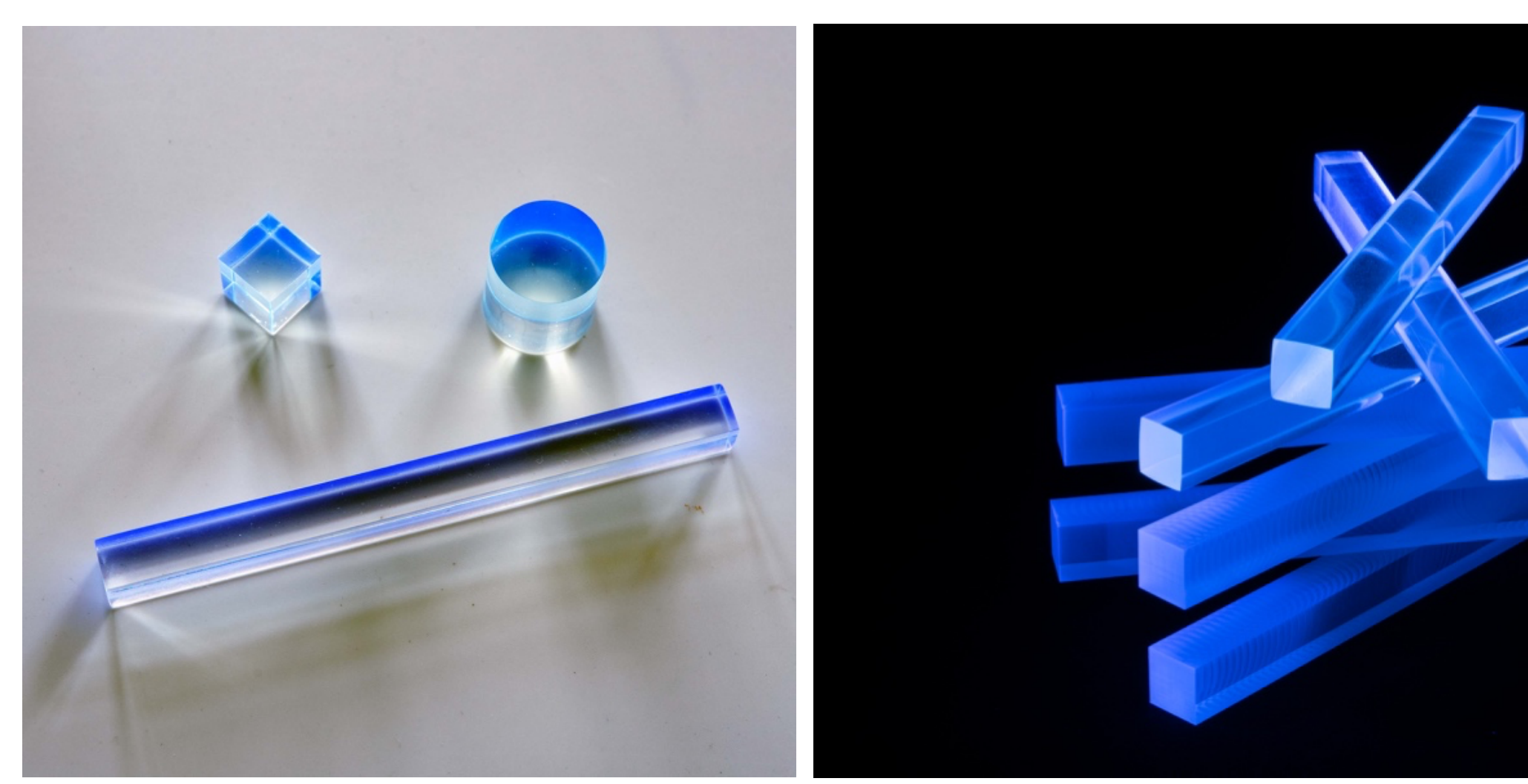

Figure 3. Examples of polyurethane-based scintillators produced by Rapiscan Systems: M700 scintillators of different shapes (left panel), and some of the $12\times12\times140$ mm$^3$ M600 bars produced for NOVO (right panel; © A. Grützner, HZDR, 2025

## 3.2 Shock-resistant radiation detector for firefighters

The easy implementation of light sensors in PU-based scintillator materials enables new ways to construct compact and shock-resistant radiation detectors. Rapiscan Systems manufactured a series of prototype detector blocks, consisting of M700 scintillators with embedded silicon photomultiplier, for ACDC-PRO[4]. This project, funded by the German Bundesministerium für Forschung, Technologie und Raumfahrt (BMFTR), aims at the development of a compact warning device for hazardous gases and ionizing radiation, designed to be worn by firefighters.

Figure 4 shows one of the detector blocks (modules) that have been produced and characterized. An M700 scintillator of about $12 \times 24 \times 40$ mm$^3$ volume comprises a single Broadcom AFBR-S4N44P014M SiPM circuit, arranged on a carrier board (PCB) together with temperature sensor, passive components, and external connectors protruding from the scintillator block (Figure 4 (b) and (d)). The inner PCB face is provided with white coating to not absorb scintillation light (Figure 4 (c)). The module was wrapped in PTFE sealing tape (reflector) and black insulation tape (Figure 4 (d)). The rugged detector construction with embedded SiPM prevents light losses due to – otherwise unavoidable – layers of optical glue or grease. It allows corresponding modules to be plugged in a motherboard, and to be replaced easily if needed.

[3] https://www.novo-project.eu

[4] https://www.sifo.de/sifo/shareddocs/Downloads/P-Umrisse/projektumriss_acdc-pro.pdf?__blob=publicationFile&v=3

For testing the detector performance, the SiPM was supplied with a bias voltage of 34 V, corresponding to an overbias of about 3 V. The detector signal was fed to a slightly modified analog frontend electronics (AFE) board of a Rapiscan Guardian 501 Handheld Radioisotope Identification Device (RID)[5] used for shaping, amplification, and digitization of the detector signal. Charge spectra obtained by integrating the digitized detector signals during exposure of the module to a $^{137}$Cs source are presented in Figure 5. The lower spectrum range (right panel) exhibits well separated peaks comprising signals with distinct numbers of fired SiPM cells. The lowest peak (just above threshold) corresponds to 4 fired cells, but even cell counts above 20 are distinguishable. This allows the direct calibration of the charge scale, reflecting the number of detected scintillation photons, in fired SiPM cells. The left panel of Figure 5 shows the characteristic Compton edge at 477 keV due to 661 keV gamma rays of the $^{137}$Cs source, which is located at about 50 SiPM cell counts. The spectrum looks surprisingly good, despite the small active area (about $3.7 \times 3.6$ mm$^2$) and a rather low overall number of 8334 SiPM cells. This compact, extremely rugged detector module is obviously not only suitable for radiation alerts or dosimetry but could even be used for gamma-ray spectroscopy, with statistical limitations given by the scintillator's light yield and the energy expense of about 10 keV per detected photon.

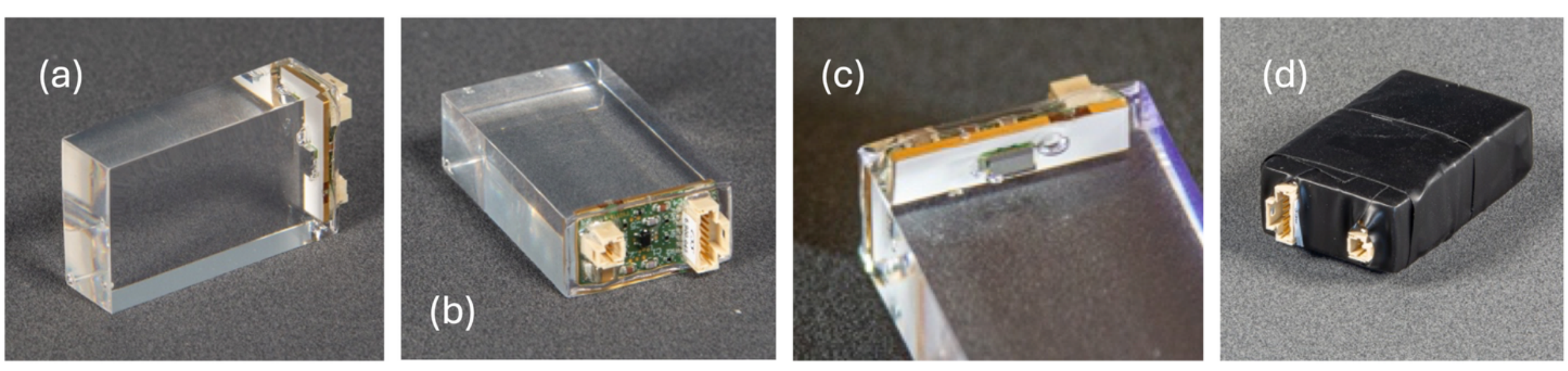

Figure 4. Photographs of an M700 scintillator with embedded light readout. This module provided the test results presented in Figure 5. (a)-(c) cured M700 casting with embedded PCB and SiPM on the white-coated board face; (d) detector block with reflector and protective wrapping. Small air bubbles due to incomplete degassing of the scintillator body during the curing process are visible in (c). Interestingly, such imperfections have only little impact on the detector performance.

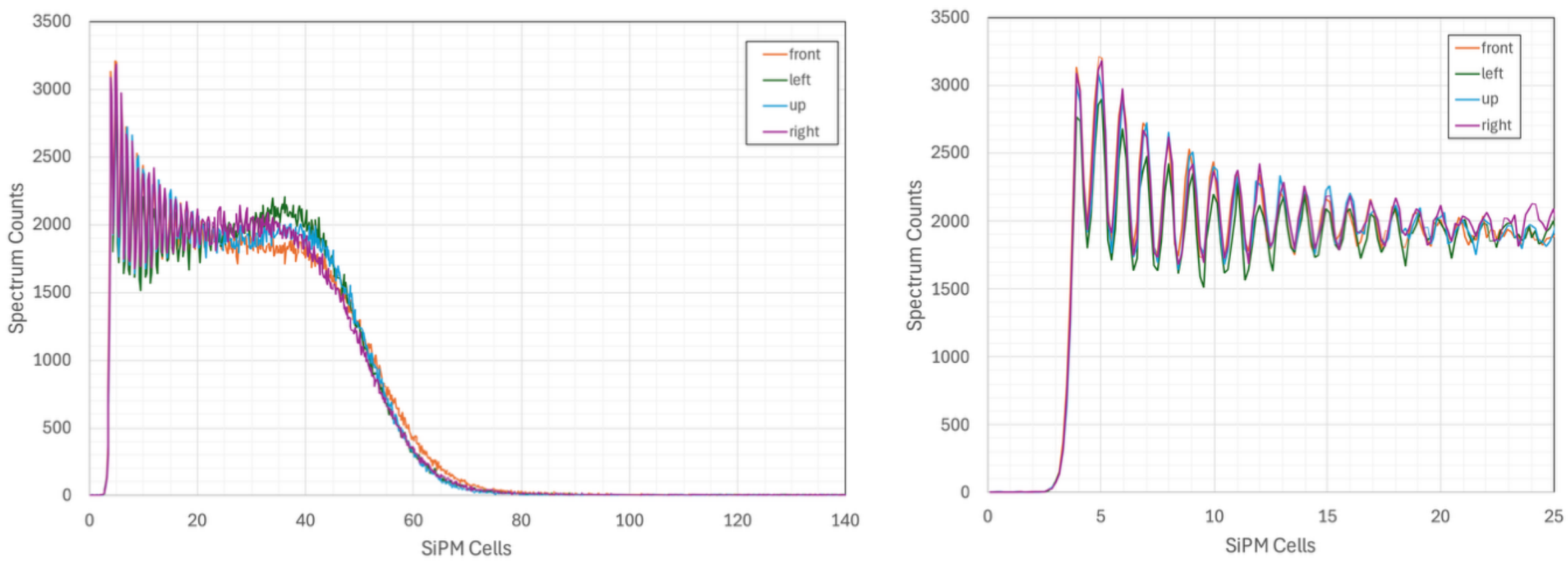

Figure 5. Energy spectra, measured with the detector shown in Figure 4, with a $^{137}$Cs source arranged in different positions (front – source on the largest block face; left / right – source near the narrow sides; front – source opposite to the SiPM). The spectrum shape is almost independent of the source position. Well resolved single-cell peaks (right panel) allow to translate spectrum channel numbers in fired SiPM cells. The position of the Compton edge at about 50 SiPM cells (left panel) corresponds to a sensitivity of about one single cell count (or single photon detection) per 10 keV of deposited energy.

[5] https://www.rapiscan-ase.com/radiation-detection-products/handheld-and-wearable/guardian-501-series

# 4. CONCLUSIONS

Organic scintillators produced by embedding a primary scintillation fluor and a wavelength shifter in a two-component polyurethane matrix have moved beyond the laboratory research stage and entered the application phase.

With respect to light yield and pulse-shape discrimination performance, corresponding materials compete with (or even outperform) the best conventional PSD plastics on the market. They combine, however, scintillation performance with the mechanical robustness and flexibility, the stress, scratch, and chemical resistance, and the long-term environmental stability of polyurethane. Another advantage is the easy production process. The components are just mixed in a liquid stage, cast in a mold, and cured at a moderate temperature around 100°C. This allows not only to produce complex scintillator shapes but also to integrate light sensors in the mold and thus in the cured scintillator.

First applications are on the way. The dual-particle imaging system NOVCoDA is expected to be constructed with M700 scintillator bars. The radiation sensor of a warning device for hazardous gases and ionizing radiation, to developed as part of the ACDC-PRO project, will be based on an M700 scintillator with integrated SiPM. The experience gained in these projects, performed in collaboration with external research groups and industrial partners, will determine which further applications are possible.